\documentclass[aps,pra,amsmath,amssymb,reprint,superscriptaddress]{revtex4-2}
\usepackage{physics}

\usepackage{graphicx}
\usepackage{dcolumn}
\usepackage{bm}

\usepackage{mathrsfs}

\usepackage{amsthm}

\usepackage{amsmath, amssymb, amsfonts, amsthm}
\usepackage{mathtools}
\usepackage{multirow}

\usepackage{color}
\usepackage{url}
\usepackage[colorlinks]{hyperref}
\usepackage{mhchem}
\hypersetup{%
        plainpages=true,
        breaklinks=true,
        hypertexnames=false,
        pageanchor=true,
        colorlinks=true,
        linkcolor={blue},
        citecolor={magenta},
        urlcolor={blue},
        anchorcolor={black}
      } 
\usepackage{mleftright}

\newcommand{\secref}[1]{\mbox{Sec.~\ref{#1}}}

\renewcommand{\eqref}[1]{\mbox{Eq.~(\ref{#1})}}
\newcommand{\be}{\begin{equation}}
\newcommand{\ee}{\end{equation}}
\newcommand{\bea}{\begin{eqnarray}}
\newcommand{\eea}{\end{eqnarray}}

\begin{document}

\preprint{APS/123-QED}

\title{Experimental detection of entanglement in multimode Gaussian states from high-order intensity correlation moments}

\author{Ze-Shan He}
\affiliation{Laboratory of Quantum Information, University of Science and Technology of China, Hefei 230026, China}
\affiliation{Hefei National Laboratory, University of Science and Technology of China, Hefei 230088, China}
\affiliation{Anhui Province Key Laboratory of Quantum Network, University of Science and Technology of China, Hefei 230026, China}
\affiliation{CAS Center for Excellence in Quantum Information and Quantum Physics, University of Science and Technology of China, Hefei 230026, China}
\author{Yukuan Zhao}
\affiliation{Laboratory of Quantum Information, University of Science and Technology of China, Hefei 230026, China}
\affiliation{Anhui Province Key Laboratory of Quantum Network, University of Science and Technology of China, Hefei 230026, China}
\affiliation{School of Physical Sciences, University of Science and Technology of China, Hefei 230026, China}
\author{Hao-Shu Tian}
\affiliation{Laboratory of Quantum Information, University of Science and Technology of China, Hefei 230026, China}
\affiliation{Hefei National Laboratory, University of Science and Technology of China, Hefei 230088, China}
\affiliation{Anhui Province Key Laboratory of Quantum Network, University of Science and Technology of China, Hefei 230026, China}
\affiliation{CAS Center for Excellence in Quantum Information and Quantum Physics, University of Science and Technology of China, Hefei 230026, China}
\author{Kai Sun}
\author{Xiao-Ye Xu}
\email{xuxiaoye@ustc.edu.cn}
\author{Chuan-Feng Li}
\email{cfli@ustc.edu.cn}
\author{Guang-Can Guo}
\affiliation{Laboratory of Quantum Information, University of Science and Technology of China, Hefei 230026, China}
\affiliation{Hefei National Laboratory, University of Science and Technology of China, Hefei 230088, China}
\affiliation{Anhui Province Key Laboratory of Quantum Network, University of Science and Technology of China, Hefei 230026, China}
\affiliation{CAS Center for Excellence in Quantum Information and Quantum Physics, University of Science and Technology of China, Hefei 230026, China}

\date{\today}


\begin{abstract}
Quantum universal invariants of a Gaussian state's covariance matrix, which can be derived from intensity correlation moments, have been adopted to characterize the entanglement properties of Gaussian states via the positive partial transpose criterion, also known as the Peres-Horodecki separability criterion. Such intensity correlation moments enable the extraction of information about the covariance matrix without the need for a coherent local oscillator. Here, we experimentally detect the entanglement properties of multimode Gaussian states using high-order\,(up to sixth-order) intensity correlation moments. These multimode Gaussian states are prepared via spontaneous and cascaded parametric down-conversion pumped by a high-peak-energy pulsed laser. Their intensity correlation moments are measured using a pseudo-photon-number-resolving detector constructed through spatial multiplexing of 32 threshold superconducting nanowire single photon detectors. This method is successfully demonstrated for two-mode and three-mode Gaussian states and can be extended to $N$-mode Gaussian states with $N>3$.

\end{abstract}

\maketitle

\section{\label{sec1}Introduction}
Entanglement, which distinguishes the quantum from the classical world, is a key resource \,\cite{nielsen2010quantum,RevModPhys.81.865}. It enables enhanced measurement precision in quantum metrology\,\cite{giovannetti2004quantum,huang2024entanglement,zhang2025entanglement} and improved security in quantum communication\,\cite{PhysRevA.95.012315,RN355}. It also underpins the advantage of quantum computation\,\cite{beth2005quantum,zhao2406entanglement,RN356}. Therefore, entanglement detection is essential for assessing whether a given quantum source is suitable for specific quantum information tasks.
Over the past decades, many feasible methods have been proposed and experimentally implemented to quantify or witness entanglement. In continuous-variable\,(CV) photonic systems, the quantum state can be fully characterized by homodyne detection\,\cite{PhysRevLett.70.1244,RevModPhys.81.299}, and the resulting characterization can be used to detect entanglement\,\cite{PhysRevLett.102.020502,yokoyama2013ultra}. 
Another effective method is based on the second-R\'enyi entropy. In the entropy criterion, entanglement is detected by comparing the entropy of the full system with that of the bipartite subsystems\,\cite{islam2015measuring,lukin2019probing}. When the entropy of the full system is smaller than that of its reduced subsystems, entanglement is present between the bipartite subsystems. 
%
However, such methods usually require a coherent local oscillator or interference among multiple copies of the quantum state. A method for detecting entanglement that reduces the need for such auxiliary resources would therefore be advantageous.

Alternatively, measuring the probability distribution in Fock space using photon-number-resolving\,(PNR) detectors has been proposed to detect the relevant quantum correlations without requiring auxiliary systems\,\cite{pevrina2020nonclassicality,barasinski2023quantification,gondret2025quantifying}. 
Native PNR detection can be implemented using superconducting transition-edge sensors\,\cite{10.1117/12.852221,FUKUDA2024135} or by exploiting the photon-number-dependent rising edge in conventional superconducting nanowire single-photon detectors\,(SNSPDs). Multiplexing threshold detectors, for example via temporal\,\cite{sperling2015uncovering} or spatial\,\cite{pevrina2014absolute} multiplexing, can be used to realize pseudo-PNR detection. 
In this work, we spatially multiplex 32 SNSPDs to construct a pseudo-PNR detector capable of fully reconstructing the correlated photon-number distribution. From this distribution, the presence of entanglement in multimode Gaussian states can be inferred via the positive partial transpose\,(PPT, also known as Peres-Horodecki separability) criterion\,\cite{HORODECKI1997333,PhysRevLett.84.2726}.
%
Furthermore, in the two-mode case, the symplectic eigenvalues that quantify entanglement can be determined explicitly, and they reveal statistical information about the partial photon number distribution.

Under the PPT criterion, the symplectic quantum universal invariants\,(QUIs) of the partially transposed covariance matrix, which we refer to simply as the symplectic QUIs $\tilde \Delta_j^N$, can be used to detect entanglement in multimode Gaussian systems, specifically for $(1+n)$-mode Gaussian states and $(m+n)$-mode bisymmetric Gaussian states\,\cite{serafini2006multimode} (here we consider the former case). For the two-mode squeezed vacuum state\,(TMSV), entanglement can be detected via the PPT criterion. The symplectic eigenvalues of the partially transposed covariance matrix, which we refer to simply as the symplectic eigenvalues $\tilde\lambda$, can be determined from intensity correlation moments\,\cite{gondret2025quantifying}. In this case, when the smallest symplectic eigenvalue is less than 1, the state is entangled, and the smaller the symplectic eigenvalue, the stronger the entanglement. For three-mode Gaussian states\,(TMGS), due to the lack of phase information, additional terms arise that cannot be expressed solely in terms of intensity correlation moments. In that case, some symplectic QUIs can be determined exactly, while others can only be bounded from above and below using intensity correlation moments. Therefore, we rely on the PPT criterion for entanglement detection in this scenario\,\cite{serafini2006multimode}.

This work is organized as follows. In \secref{sec2}, we introduce the theoretical framework for deriving the symplectic QUIs of the TMSV and TMGS from intensity correlation moments. We further derive the symplectic eigenvalues of the TMSV and provide the corresponding condition for entanglement detection. In \secref{sec3}, we describe the experimental preparation of the TMSV and TMGS. The TMSV is produced via spontaneous parametric down-conversion\,(SPDC), in which a pulsed pump laser is focused into a customized quasi-phase-matched \ce{KTiOPO4}\,(cpKTP) crystal. The TMGS is generated via cascaded stimulated parametric down-conversion, in which one mode of the TMSV is sent into a second cpKTP crystal simultaneously with the pump laser from the original source. Finally, we present experimental and simulated results demonstrating violation of the PPT criterion for the two Gaussian states, thereby confirming the presence of entanglement in our system. We conclude with a summary in \secref{sec4}.

\section{\label{sec2}Theoretical background}
For an $N$-mode Gaussian state\,(specifically, $(1+n)$-mode Gaussian states and $(m+n)$-mode bisymmetric Gaussian states), the PPT criterion yields the following necessary condition for separability: $\tilde \sigma+i\Omega \geq0$\,\cite{serafini2006multimode}, where $\tilde \sigma$ denotes the partially transposed covariance matrix and $\Omega = \bigoplus_{k=1}^N is_y$ with $s_y$ the Pauli-$y$ matrix. This condition can be rewritten as
\begin{equation}
    \sum_{k=0}^{N}(-1)^{N +k} \tilde \Delta _k^N\geq0.
    \label{eq:ppt}
\end{equation}
The symplectic QUIs of the partially transposed covariance matrix $\tilde \Delta_k^N$ are defined as the principal minors of order $2k$ of the matrix $\Omega \tilde\sigma$.

In this section, we present the theoretical framework for obtaining symplectic QUIs and symplectic eigenvalues for the TMSV, as well as the symplectic QUIs of the TMGS, from intensity correlation moments\,\cite{barasinski2023quantification,gondret2025quantifying,sudak2025efficient}. For the TMSV, we use Wick's theorem\,\cite{kardar2007statistical} to obtain the correlation functions, thereby determining both symplectic QUIs and corresponding symplectic eigenvalues explicitly. For the TMGS, we express the state parameters in terms of intensity correlation moments. Owing to the absence of phase information or a reference light, only a subset of the symplectic QUIs can be obtained. Nevertheless, this information is sufficient to assess entanglement via the PPT criterion.

\subsection{\label{sec2:A} Symplectic QUIs and symplectic eigenvalues of the TMSV}
For the TMSV,~\eqref{eq:ppt} can be simplified to
\begin{equation}
    \tilde \Delta^2_2+1\geq \tilde \Delta_1^2.
    \label{eq:ppt_2mode}
\end{equation}
This conditions is also equivalent to $\tilde \lambda_- \geq 1$, where $\tilde \lambda_-$ denotes the smallest symplectic eigenvalue of the TMSV. The symplectic eigenvalues can be expressed as
\begin{equation}
    \tilde{\lambda}_{\pm}^2=\frac{\tilde \Delta_1^2\pm\sqrt{(\tilde \Delta_1^2)^2-\tilde \Delta^2_2}}{2}.
    \label{Symplectic_eigenvalues_quis}
\end{equation}

The two symplectic QUIs and symplectic eigenvalues of the TMSV can be obtained from intensity correlation moments without the need for a phase reference. This approach shows promise for extension to $N$-mode Gaussian states with $N>3$. 

The covariance matrix of the TMSV is defined as $\sigma \equiv \langle \hat{\xi}\hat{\xi}^T \rangle -\langle \hat{\xi} \rangle \langle \hat{\xi}^T \rangle $, where $ \langle \hat{\xi}^T \rangle =\left (\hat{x}_1,\hat{p}_1,\hat{x}_2,\hat{p}_2 \right)$ is the vector of quadrature operators, and $\hat{x}_i $, $\hat{p}_i$ are the phase-space operators of the Gaussian state. This covariance matrix can be written in block form\,\cite{barasinski2023quantification}.

\begin{equation}
    \begin{aligned}
    \sigma =
    \begin{bmatrix}
    \sigma_1 & \gamma_{12} \\
    \gamma_{12}^T & \sigma_2
    \end{bmatrix},
    \end{aligned}
    \label{eq:cm_tb}
\end{equation}
where $\sigma_j$ denotes the $2\times 2$ local covariance matrices and
$\gamma_{jk}$ denotes the $2\times 2$ intermode correlation block matrices. Using the normal-ordered characteristic function\,\cite{barasinski2023quantification}, $\sigma_j$ and $\gamma_{ij}$ can be expressed as
\begin{equation}
    \begin{aligned}
    \sigma_j &=
    \begin{bmatrix}
    1 + 2B_j + 2\Re C_j & 2\Im C_j \\
    2\Im C_j & 1 + 2B_j - 2\Re C_j
    \end{bmatrix}, \\
    \gamma_{jk} &=
    \begin{bmatrix}
    2\Re(D_{jk}-\bar D_{jk}) & 2\Im(D_{jk}-\bar D_{jk}) \\
    2\Im(D_{jk}+\bar D_{jk}) & -2\Re(D_{jk}+\bar D_{jk})
    \end{bmatrix}.
    \end{aligned}
    \label{block_matrix}
\end{equation}

The four parameters, defined as $B_j=\langle \hat{a}_j^\dagger \hat{a}_j \rangle$, $C_j=\langle \hat{a}_j\hat{a}_j \rangle$, $D_{jk}=\langle \hat{a}_j \hat{a}_k \rangle, -\bar{D}_{jk}^{\ast}=\langle \hat{a}_j \hat{a}_k^\dagger\rangle$ for $(j\neq k)$, can be determined from intensity correlation moments in~\eqref{BCD_moments}, and the details of other terms are provided in the supplementary material.
\begin{equation}
    \begin{aligned}
    B_j&=\langle W_j\rangle \\
    \left|C_j\right|^2 &= \langle W_j^2\rangle-2\langle W_j\rangle ^2,\quad j=1,2, \\
    \left|D_{ij}\right|^2 + \left|\bar{D}_{ij}\right|^2&= \langle W_iW_j\rangle -\langle W_i\rangle\langle W_j\rangle.
    \label{BCD_moments}
    \end{aligned}
\end{equation}

The partially transposed symplectic eigenvalue depends on the two partially transposed symplectic QUIs: the first, $\tilde \Delta_1^2$, is determined by the determinants of the local block matrices $\sigma_j$ and the intermode block matrix $\gamma_{12}$, while the second, $\tilde \Delta_2^2$, is determined by the determinant of the global covariance matrix $\sigma$.

The first symplectic QUI can be written as $\tilde \Delta_1^2 =\det\sigma_1+\det\sigma_2 - 2\det\gamma_{12}$. Combining~\eqref{block_matrix} and~\eqref{BCD_moments}, we obtain $\det\sigma_j=1+4\langle W_j\rangle+12\langle W_j\rangle^2-4\langle W_j^2\rangle$ and $\det\gamma_{12}=-4(|D_{12}|^2-|\bar{D}_{12}|^2)$. The former can be obtained from intensity correlation moments, while the latter can be evaluated using $g^2_{12}$ and $g^4_{12}$\,\cite{gondret2025quantifying} between the two modes. This follows from the fact that the TMSV satisfies $\langle \hat{a}_i\rangle=0$ and  $\langle \hat{a}_i^2\rangle=0$, implying that each reduced mode exhibits a thermal state photon-number distribution. Using Wick's theorem $\langle \hat{a}_i \hat{a}_j \hat{a}_k \hat{a}_l \rangle = \langle \hat{a}_i \hat{a}_j \rangle\langle \hat{a}_k \hat{a}_l \rangle + \langle \hat{a}_i \hat{a}_k \rangle\langle \hat{a}_j \hat{a}_l \rangle+\langle \hat{a}_i \hat{a}_l \rangle\langle \hat{a}_j \hat{a}_k \rangle$, we obtain
\begin{equation}
    \begin{aligned} 
    g^{(2)}_{12} :&=\frac{ \langle \hat{a}^\dagger_1 \hat{a}^\dagger_2 \hat{a}_1 \hat{a}_2 \rangle}{n_1 n_2} \\
    &=\frac{\langle \hat{a}^\dagger_1 \hat{a}^\dagger_2\rangle \langle \hat{a}_1 \hat{a}_2 \rangle+\langle \hat{a}^\dagger_1 \hat{a}_1\rangle \langle \hat{a}^\dagger_2 \hat{a}_2 \rangle+\langle \hat{a}^\dagger_1 \hat{a}_2\rangle \langle \hat{a}^\dagger_2 \hat{a}_1 \rangle}{n_1n_2}\\
    &=1+ \frac{\left |\langle \hat{a}_1 \hat{a}_2 \rangle \right |^2}{n_1 n_2} +\frac{\left |\langle \hat{a}_1 \hat{a}^\dagger_2 \rangle \right |^2}{n_1 n_2}
    \label{g_2}
    \end{aligned}
\end{equation}
and
\begin{equation}
    \begin{aligned}
    g^{(4)}_{12}  :&=\frac{ \langle (\hat{a}^\dagger_1 \hat{a}^\dagger_2 )^2 (\hat{a}_1 \hat{a}_2)^2 \rangle}{n_1^2 n_2^2} \\
    &=4\left[1+\left( g^{(2)}_{12}-1\right)^2+4 \left  (g^{(2)}_{12}-1 \right ) \right.\\
    &\quad \left. +2\frac{|\langle \hat{a}_1 \hat{a}^\dagger_2  \rangle |^2 |\langle \hat{a}_1 \hat{a}_2  \rangle|^2}{n_1^2 n_2^2}\right],
    \label{g_4}
    \end{aligned}
\end{equation}
where $n_1$ and $n_2$ are the mean photon numbers of the two modes of the TMSV. Therefore, the squared moduli $|D_{12}|$ and $|\bar{D}_{12}|$ can be obtained by combining the two equations above, which is essential for determining the symplectic eigenvalues of the TMSV. To simplify the notation, we define $\beta_+ = |D_{12}|$ and $\beta_-=|\bar{D}_{12}|$. Solving these two equations yields
\begin{equation}
    \beta^2_{\pm}=n_1 n_2 (g^{(2)}_{12}-1)\frac{1\pm \sqrt{1-\theta}}{2},
    \label{beta}
\end{equation}
where $\theta $ is given by
\begin{equation}
    \theta = \frac{g^{(4)}_{12}+12-16g^{(2)}_{12}-4\left(g^{(2)}_{12}-1\right)^2}{2\left(g^{(2)}_{12}-1\right)^2}.
    \label{theta}
\end{equation}

Thus, the first symplectic QUI can be expressed as  
\begin{equation}
    \begin{aligned}
    \tilde\Delta_1^2 =& 2+\{4\langle W_1\rangle+12\langle W_1\rangle^2-4\langle W_1^2\rangle+\mathrm{addt1}\}\\
    &+8(\beta^2_+-\beta^2_-),
    \label{delta_1}
    \end{aligned}
\end{equation}
where the symbol $\mathrm{addt1}$ denotes the terms obtained by the substitutions $\{1\}\to\{2\}$. 

The second symplectic QUI $\tilde\Delta_2^2$ is given by the determinant of the global covariance matrix $\sigma$\,\cite{serafini2003symplectic}. By combining~\eqref{block_matrix} and~\eqref{BCD_moments}, this QUI can be expressed in terms of intensity correlation moments as $\det\sigma  =f\left[\langle W_i\rangle,\langle W_iW_j\rangle,\langle W_i^2W_j\rangle,\langle W_i^2W_j^2\rangle\right]$. The explicit expressions are provided in the supplementary material. From these quantities, the smallest symplectic eigenvalue can be obtained, as given by~\eqref{Symplectic_eigenvalues_quis}.

As mentioned earlier, violation of the condition in~\eqref{eq:ppt_2mode} implies that the state is entangled. Equivalently, the state is entangled when $\tilde\lambda_-<1$, and separable when $\tilde \lambda_-\geq1$. Thus, by using intensity correlation moments, the symplectic eigenvalues can be obtained, enabling entanglement detection.
%


\subsection{\label{sec2:B} Symplectic QUIs of the TMGS}
For the asymmetric TMGS, the symplectic eigenvalues cannot be directly derived from intensity correlation moments due to the lack of phase information or reference light. Nevertheless, the accessible moments still determine part of the symplectic QUIs, while upper and lower bounds can be established for the remaining phase-sensitive terms. This enables a bounded formulation of the PPT criterion and thereby allows entanglement to be assessed.

The symplectic QUIs for the TMGS are more involved, and there are three such symplectic QUIs. The covariance matrix of the TMGS takes the form

\begin{equation}
    \begin{aligned}
    \sigma =
    \begin{bmatrix}
    \sigma_1 & \gamma_{12} & \gamma_{13} \\
    \gamma_{12}^T & \sigma_2 & \gamma_{23} \\
    \gamma_{13}^T & \gamma_{23}^T & \sigma_3 
    \end{bmatrix},
    \end{aligned}
    \label{convariance_matrix_for_three}
\end{equation}
where $\sigma_i$ and $\gamma_{ij}$ are $2 \times 2$ block matrices with $i,j \in \{1,2,3\}$. We first introduce the real symplectic QUIs $\Delta^N_j $ of the TMGS\,(which will later be used to derive the symplectic QUIs of the partially transposed covariance matrix $\tilde\Delta^N_j$, the details are given in the supplementary material).
The first symplectic QUI can be expressed as
\begin{equation}
    \begin{aligned}
    \Delta^3_1 &=3+\{4\langle W_1\rangle +12\langle W_1\rangle ^2-4\langle W_1^2\rangle +\text{addt1}\}-\Delta^3_{1,\beta}\\
    &=\Delta^3_{1,w}-\Delta^3_{1,\beta}.
    \end{aligned}
    \label{first_symplectic_invariants}
\end{equation}
Here $\Delta^3_{1,\beta}=\{4|D_{12}|^2-|\bar{D}_{12}|^2+\mathrm{addt2}\}$, whose value can also be obtained from~\eqref{beta}, provided that the state satisfies $\langle \hat{a}_i\rangle=0$ and  $\langle \hat{a}_i^2\rangle=0$. Here $\mathrm{addt1}$ denotes the terms obtained by replacing the index 1 with indices 2 and 3, respectively, and $\mathrm{addt2}$ denotes the sum over all distinct pairs of indices chosen from $\{1,2,3\}$.  We note that the first symplectic QUI of the TMGS is similar to that in~\eqref{delta_1}, because it is given by a sum of determinants of submatrices of the covariance matrix. The second symplectic QUI can be expressed as
\begin{equation}
    \begin{aligned}
    \Delta^3_2 =\Delta^3_{2,w}-\Delta^3_{2,r},
    \end{aligned}
    \label{second_symplectic_invariants}
\end{equation}
$\Delta^3_{2,w}$ denotes the term can be obtained from intensity correlation moments, while $\Delta^3_{2,r}$ denotes the residual term in the second symplectic QUI\,\cite{sudak2025efficient}. As mentioned above, due to the lack of phase information, several contributions to the second QUI, such as $\rm{Re}\{C_1\bar{D}_{12}\bar{D}_{13}D^\ast_{23}\}$, cannot be directly obtained from intensity correlation moments\,(see the supplementary material for further details).
The third symplectic QUI is given by the determinant of the covariance matrix and can be expressed as $\Delta_3^3=\det\sigma$\,\cite{sudak2025efficient}.

In fact, the PPT criterion implies that the TMGS is separable if it satisfies the inequality $\tilde{\Delta}_3^3-\tilde{\Delta}_2^3+\tilde{\Delta}_1^3-1\geq0$ from~\eqref{eq:ppt}, where $\tilde{\Delta}_k^N$ denotes the symplectic QUI of the partially transposed covariance matrix.
The symplectic QUIs of the partially transposed covariance matrix with respect to the first mode can be obtained from the original symplectic QUIs by applying the partial-transposition transformation.
\begin{equation}
    \begin{aligned}
    \tilde{\Delta}_1^3&=\Delta^3_{1,w}-\tilde\Delta^3_{1,\beta} \\
    \tilde{\Delta}_2^3&=\Delta^3_{2,w}-\tilde\Delta^3_{2,r}\\
    \tilde{\Delta}_3^3&=\Delta^3_3.
    \end{aligned}
    \label{symplectic_invariants_partial_transposition}
\end{equation}

In this case, the PPT criterion can be expressed as
\begin{equation}
 \label{symply_symplectic_invariants_inequality}
\begin{aligned}
    &&\Delta^{3}_{3}-\Delta^3_{2,w}
    +\Delta_1^3 + 2 \Delta_{1,\beta}^3 - 16r_{23} -1\\&&+16\bigl(|\bar D_{12}|^2 + |\bar D_{13}|^2 + |D_{23}|^2\bigr)\\
    &&\ge
    -\sum_{\pi\in S_3} \tilde F[\pi(1),\pi(2),\pi(3)],
\end{aligned}
\end{equation}
with $r_{jk} = |D_{jk}|^2 - |\bar{D}_{jk}|^2$, and $\tilde{F}$ denoting all the residual terms\, see the supplement material for more details. The left-hand side\,(LHS) of the inequality contains terms that can be expressed in terms of intensity correlation moments, while the remaining terms appear on the right-hand side\,(RHS). Upper and lower bounds for the RHS can be obtained as follows:

\begin{equation}
\begin{aligned}
|-\tilde F_{123}|
&\le 16\left|B_1^{2}+B_1- |C_1|^{2}\right|\,\Sigma_{23}
   + 16\,\Sigma_{12}\Sigma_{13} \\
&\quad + \sqrt{\Sigma_{12}\Sigma_{13}\Sigma_{23}}
      \biggl[16 + 32|B_1 + B_2 - B_3| \\
&\quad + 16(2B_1+1) + 64\sqrt{\langle W_1^{2}\rangle - 2\langle W_1\rangle^{2}} \\
&\quad + 64\sqrt{\langle W_2^{2}\rangle - 2\langle W_2\rangle^{2}}\biggr],
\end{aligned}
\label{eq:F123-tight-intensity}
\end{equation}

where $\Sigma_{ij}=|D_{ij}|^2+|\bar{D}_{ij}|^2$. A similar inequality holds for other permutations, corresponding to $\tilde{F}[\pi(1),\pi(2),\pi(3)]$. According to the PPT criterion, if the LHS of~\eqref{symply_symplectic_invariants_inequality} is smaller than the lower bound of the RHS, the state is entangled. We demonstrate this result by measuring a TMGS with a 32-channel PNR detector, as described in Sec.\,\ref{sec3}.


\section{\label{sec3}Experimental setup and results}
We prepare the TMSV and TMGS via spontaneous and stimulated parametric down-conversion, respectively. For each state, the intensity correlation moments among its modes are measured using a 32-channel PNR detector, from which the corresponding symplectic QUIs are extracted. The observed distortions arise from the finite number of detection channels and the multimode (time-domain supermode) structure of the pump laser. We perform theoretical simulations and compare them with the experimental data for the two Gaussian states. The experimental results are in good agreement with the simulations, confirming the presence of entanglement in both the TMSV and the TMGS.

\subsection{\label{sec3:A}Setup}
The TMSV is an ideal two-mode Gaussian state in which each mode exhibits a thermal distribution. The photon numbers in its signal and idler modes are perfectly correlated: a projective photon-number measurement on one mode determines the photon number of the other, reflecting strong entanglement in the Fock basis. We generate spectrally pure biphotons using a cpKTP crystal\,\cite{PhysRevA.93.013801,PhysRevApplied.12.034059}, such that frequency entanglement between the two modes is negligible. To characterize the generated biphotons, we measure the joint spectral intensity\,(JSI) using two 4.5\,km dispersion-compensating fibers\,(DCF, model DM1012-A), which map the spectrum from the frequency domain to the time domain. The spectral purity is then extracted from the measured JSI. The measured spectral purity is 0.9945, indicating that the biphotons state is nearly a pure state in the frequency domain. Further details are provided in the supplementary material.

In the spontaneous parametric down-conversion process, the interaction Hamiltonian that generates the TMSV is given by $\text{H}_1=i\hbar(\xi_1^{\ast}\hat{a}_1\hat{b}-\xi_1\hat{a}_1^{\dagger}\hat{b}^{\dagger})$, where $\xi_1=r_1e^{i\psi}$ is the squeezing parameter, whose magnitude is determined by the pump power, crystal length, and beam waist. The corresponding unitary operator can be written as $\text{S}(\xi_1)=\exp(\xi_1^{\ast}\hat{a}_1\hat{b}-\xi_1\hat{a}^{\dagger}_1\hat{b}^{\dagger})$\,\cite{agarwal2012quantum}. Acting on the vacuum state, this operator generates the TMSV, which can be expressed in the Fock basis as
\begin{equation}
    \begin{aligned}
    \ket{\xi_1} =& \text{S}(\xi_1)\ket{0,0}\\
     =&\frac{1}{\cosh r}\sum_{n=0}^{\infty}e^{in\psi}\tanh^nr\ket{n,n}
    \end{aligned}
    \label{TMSV}
\end{equation}

Due to the impurity of the pump laser, i.e., the presence of multimode components, we adopt a simplified multimode squeezing model\,\cite{christ2011probing}, in which the state is described as a tensor product of independent TMSVs. Accordingly, the resulting state can be written as
\begin{equation}
    \begin{aligned}
    \ket{\xi_1} =& \mathop{\bigotimes}\limits_{k}\hat{S}_k^{ab}(\xi_{1,k})\ket{0_k,0_k}.
    \end{aligned}
    \label{multi_TMSV}
\end{equation}

Here the subscript $k$ labels the supermode of the pump laser, while the subscript $1$ denotes the first squeezing process. The amplitude of the squeezing parameter is given by $r_{1,k}=r_1\sqrt{1-\mu^2}\mu^k$. The parameter $\mu$ is inferred from the second-order correlation function $g^2(0)$ via $\mu = \sqrt{\frac{2}{g^2(0)}-1}$.

Unlike a TMSV, which is generated via spontaneous parametric down-conversion, the TMGS requires cascaded stimulated parametric down-conversion. This method can be extended to generate $N$-mode Gaussian states with $N>3$\,\cite{dhand2018proposal}. Additionally, it can be combined with a projective measurement to prepare non-Gaussian states.\,\cite{jing2025generation}. The interaction Hamiltonian for the stimulated process in the second crystal is given by $\mathrm{H_2}=i\hbar(\xi_2^{\ast}\hat{a}_2\hat{b}-\xi_2\hat{a}_2^{\dagger}\hat{b}^{\dagger})$, where $\hat{b}$ denotes the signal mode of the TMSV, and $\hat{a}_2$ represents a newly generated idler mode. The resulting TMGS can then be expressed in the Fock basis as

\begin{equation}
    \begin{aligned}
    \ket{\xi_2} =& S(\xi_2)S(\xi_1)\ket{0,0,0}\\
    =& \exp(\xi_2^{\ast}\hat{a}_2^{\dagger}\hat{b}^{\dagger}-\xi_2\hat{a}_2\hat{b})\exp(\xi_1^{\ast}\hat{a}_1\hat{b}-\xi_1\hat{a}_1^{\dagger}\hat{b}^{\dagger})\ket{0,0,0}\\
    =&\frac{1}{\cosh^2r}\sum_{n_0,n_1}^{\infty}C_{n_0,n_1,n_0+n_1}\ket{n_0,n_1,n_0+n_1}.
    \end{aligned}
    \label{tbgs}
\end{equation}

The coefficient $\mathrm{C}_{n_0,n_1,n_0+n_1}$ depends on the squeezing parameter $\xi_1$ and $\xi_2$ and the photon number $n_0$ and $n_1$. The photon number in the third mode is constrained to equal the sum of those in the first and second modes, reflecting strong correlations among the three modes in the Fock basis. Taking into account the effect of pump impurity, ~\eqref{tbgs} can be rewritten in a form analogous to~\eqref{multi_TMSV}.

\begin{figure}[h]
    \centering
    \includegraphics[width=0.45\textwidth]{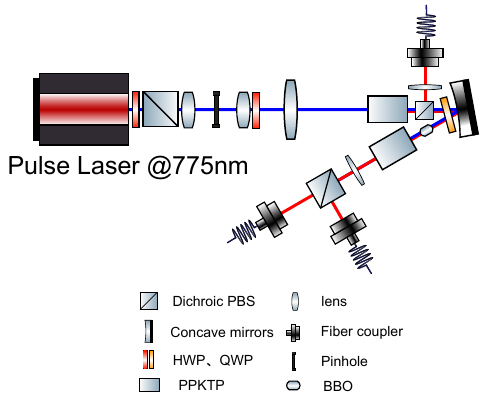}
    \caption{Generation of two-mode and three-mode Gaussian states.
    The laser from a Coherent RegA 9000 is filtered by a 4-f spatial filtering system. We use a single cpKTP crystal to generate a two-mode Gaussian state and subsequently employ two cpKTP crystals to produce a three-mode Gaussian state. The SPDC process in the first crystal generates a two-mode squeezed vacuum state: the idler photon is reflected by a dichroic PBS, while the signal photon is injected into the second crystal simultaneously with the pump laser. After the signal mode polarization is rotated by $90^\circ$, a BBO crystal compensates the time difference between the pump laser and the signal mode after polarization rotation $90^{\circ}$ of signal mode; subsequently, the stimulated process occurs in the second cpKTP crystal and thus generates a three-mode Gaussian state. All these modes are detected by PNR detectors, enabling the measurement of photon-number correlation statistics.}
    \label{threemode}
\end{figure}

The experimental setup for generating the TMSV and TMGS is shown in Fig.\,(\ref{threemode}). We use a Coherent RegA 9000 to generate the pulsed laser. It delivers high-energy pulses, up to 4\,$\mu \text{J}$ per pulse, enabling a sufficiently large squeezing parameter. The pulsed laser operates at 775\,nm, with a spectral bandwidth of 5.1\,nm and a repetition rate of 250\,kHz. Before entering the crystal, the pump laser is reshaped into the fundamental Laguerre-Gaussian\,(LG) mode in the spatial domain using a 4-f spatial filtering system. The pump laser is then focused onto the center of a cpKTP crystal using a lens with a focal length of 200\,mm, thereby generating the TMSV. 

The TMGS is generated by using one mode of the TMSV as the seed for the stimulated process. The idler mode of the TMSV is reflected by a dichroic polarizing beam splitter\,(PBS), while the signal mode is transmitted together with the pump laser.  
In the stimulated process, the seed and pump arrive simultaneously at the center of the second cpKTP crystal. The first and third modes are then entangled with the second mode via the stimulated process. To achieve this, the polarization of the signal mode is rotated by $90^{\circ}$ using a double pass through a quarter-wave plate (QWP), while the pump polarization remains unchanged. A BBO crystal is then used to compensate for the \textcolor{red}{300\,fs} delay introduced by the dichroic PBS, ensuring temporal synchronization between the pump laser and the signal mode.

We use three fiber couplers\,(60FC-SF-0-A15-45) to collect the photons from the three modes. Photons from two of the modes are combined into a single spatial mode using an electro-optic modulator\,(EOM), while the third mode is kept separate, as the 32-channel PNR detector has two input ports. A fiber‑induced time separation of 180\,ns is maintained between successive modes, realizing a time‑division multiplexing scheme that allows all three modes to be detected by the same PNR detector. The two spatial modes are then sent to the detector, which has a total collection efficiency of $82\%$\,(including coupling and detection losses).


\subsection{\label{sec3:B}Results}
\begin{figure}[htpb]
    \centering
    \includegraphics[width=0.45\textwidth]{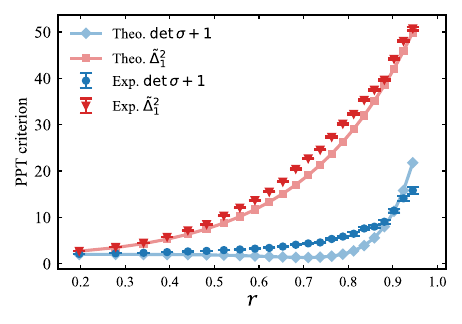}
    \caption{PPT criterion inequality for a TMSV. The experimental and simulated results for the inequality~\eqref{eq:ppt_2mode} as a function of $r$ for $g^2(0)=1.8$ are shown. The light red solid line with square markers and the light blue solid line with diamond markers represent the simulated values of the two sides of the inequality, corresponding to $\det\sigma+1$ and $\tilde\Delta_1^2$, respectively. The blue and red dots represent the experimental results, which are in good agreement with the simulations. Specifically, the blue dots correspond to $\det\sigma+1$ and the red dots correspond to $\tilde \Delta_1^2$, with $r$ ranging from $0.18$ to $0.96$.}
    \label{ppt_two_mode}
\end{figure}

\begin{figure}[htpb]
    \centering
    \includegraphics[width=0.45\textwidth]{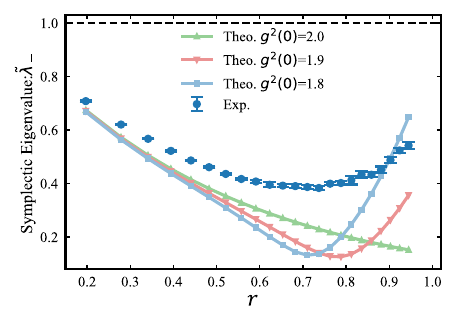}
    \caption{Symplectic eigenvalues of the TMSV as a function of $r$. Experimental and simulated results for the smallest symplectic eigenvalue of the partially transposed covariance matrix $\tilde\lambda_-$ are shown. The green solid line with up-triangle marker corresponds to $g^2(0)=2$, i.e., a pure TMSV. The red and blue solid lines correspond to $g^2(0)=1.9$ and $g^2(0)=1.8$, respectively; the latter was obtained using experimentally measured parameters with a PNR detector. The blue dots represent the experimental results. For $r>0.75$, the eigenvalue increases due to the growing contribution of additional pump supermodes, which modifies the photon-number distribution. The black dashed line marks the threshold $\tilde\lambda_-=1$; values below this line correspond to the presence of entanglement.}
    \label{sympletic_eigenvalue_two_mode}
\end{figure}

In the case of the TMSV, we use~\eqref{eq:ppt_2mode} and~\eqref{multi_TMSV} to simulate the PPT inequality as a function of $r$, and then extract the symplectic eigenvalues from the two symplectic QUIs. The simulated and experimental results are shown in Fig.\,(\ref{ppt_two_mode}) and Fig.\,(\ref{sympletic_eigenvalue_two_mode}), respectively.

As mentioned above, an observed value of $g^2(0)<2$ indicates the presence of multiple modes\,(i.e., supermodes) in the pump laser. 
In Fig.\,\ref{ppt_two_mode}, we plot the simulated inequality~\eqref{eq:ppt_2mode} together with the corresponding experimental results as a function of $r$ for $g^2(0)=1.8$. In both cases, the inequality~\eqref{eq:ppt_2mode} is violated, indicating entanglement between the two modes of the TMSV across the explored parameter range. The vertical separation between $\tilde{\Delta}_1^2$ and $\det \sigma+1$ defines the experimental margin by which entanglement is certified. At small $r$, this margin is small, making the witness more sensitive to noise and detection imperfections, whereas at larger $r$ it increases, leading to more robust entanglement certification.


The ideal $g^2(0)$ value of 2 for a single mode of the TMSV originates from its thermal photon-number distribution. Here, we simulated the symplectic eigenvalues as a function of $r$ at different values of $g^2(0)$. As shown by the light green solid line in Fig.\,\ref{sympletic_eigenvalue_two_mode}, for $g^2(0)=2$, each mode is thermal and the pairwise correlations generated by the squeezing strengthen as $r$ increases. Consequently, $\tilde{\lambda}_{-}$ decreases monotonically with increasing $r$ and remains below 1, indicating the presence of entanglement between the signal and idler modes of the TMSV. In the nonideal case $g^2(0)<2$, the behavior is qualitatively different. The light red and light blue solid lines correspond to $g^2(0)=1.9$ and $g^2(0)=1.8$, respectively. In these cases, the symplectic eigenvalues initially decrease with increasing $r$, but then increase for large $r$; moreover, the smaller $g^2(0)$ is, the earlier this turnover occurs.
This behavior arises because the pump laser contains multiple supermodes, so the generated state is effectively a mixture of independently squeezed mode pairs rather than a single pure TMSV. As $r$ increases, these additional supermodes contribute more strongly to higher-order moments, thereby degrading the effective two-mode entanglement probed by the phase-insensitive witness and causing $\tilde{\lambda}_{-}$ to increase at larger squeezing.
Hence, improving the pump purity would suppress the unwanted supermodes and enable entanglement certification even at low squeezing.

The experimental results are shown as blue dots in Fig.\,\ref{sympletic_eigenvalue_two_mode}, corresponding to measurements taken at $g^2(0)=1.8$ using a 32-channel PNR detector at low pump power. The condition $\tilde \lambda_-<1$ indicates the presence of entanglement between the signal and idler modes. The discrepancy between the experimental and simulated results may be attributed to detection imperfections and optical path losses. These results are obtained solely from intensity correlation moments, demonstrating that the entanglement properties of a TMSV can be verified without phase information. 

For the TMGS, the symplectic eigenvalues cannot be directly reconstructed from intensity correlation moments due to the absence of phase-sensitive terms, as discussed above. However, upper and lower bounds for these terms can be expressed in terms of intensity correlation moments, enabling the PPT inequality to be evaluated without phase information for entanglement detection. After deriving~\eqref{eq:F123-tight-intensity}, one obtains upper and lower bounds on $\mathop{\sum}\limits_{\pi \in S_3} \tilde F(\pi(1),\pi(2),\pi(3))$. Substituting these bounds into~\eqref{symply_symplectic_invariants_inequality} allows one to assess entanglement. In particular, whenever the LHS of~\eqref{symply_symplectic_invariants_inequality} falls below the lower bound of the RHS, the Gaussian state is certified as entangled.

\begin{figure}[h]
    \centering
    \includegraphics[width=0.45\textwidth]{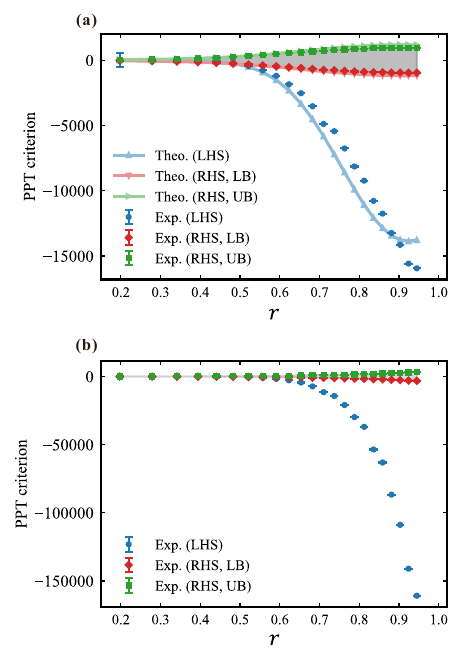}
    \caption{PPT criterion inequality for a TMGS. (a) Experimental and simulated results for inequality~\eqref{symply_symplectic_invariants_inequality} as a function of $r$ at $g^2(0)=1.8$ with a photon-number cutoff of 5. The light blue solid line with up-triangle markers represents the simulated value of the left-hand side\,(LHS) of the inequality. The light green and light red solid lines with right- and down-triangle markers represent the upper bounds\,(UB) and lower bounds\,(LB) of the right-hand side\,(RHS), respectively. Experimental results are shown as corresponding dots\,(see legend) and agree well with the simulations. (b) Experimental results for the same inequality with a photon-number cutoff of 32. The observed trend is consistent with that obtained for a cutoff of 5.}
    \label{inequality_three_mode}
\end{figure}

We numerically calculate the theoretical bounds at $g^2(0)=1.8$. Fig.\ref{inequality_three_mode}(a) shows the numerical simulation and experimental results with a photon-number cutoff of 5. When the three lines are close at small $r$, the result is inconclusive and does not constitute evidence of separability. Because the phase-sensitive residual terms cannot be directly reconstructed from intensity correlation moments, only upper and lower bounds of the RHS of the PPT inequality are known. Therefore, when the LHS lies between the bounds, the measurement does not yet certify entanglement, nor does it demonstrate the absence of entanglement. Only when the LHS of the inequality falls below the lower bound does one obtain a phase-insensitive, device-limited but rigorous certification of entanglement. The gap between the upper and lower bounds represents the cost of lacking phase information: it quantifies the uncertainty arising from unmeasured phase-sensitive contributions to the TMGS symplectic QUIs. As $r$ increases, the contribution extracted from intensity correlation moments becomes large enough that, even after allowing for the worst-case residual terms, the PPT inequality is still violated. 

Fig.\ref{inequality_three_mode}(b) shows the experimental results for the same inequality with a photon-number cutoff of 32. Due to the computational cost, simulations cannot be performed at a photon‑number cutoff of 32. Results for other photon‑number cutoffs, from both experiment and simulations, are provided in the supplementary material. The experimental results obtained at a photon‑number cutoff of 32 remain consistent with the trends observed at a cutoff of 5, indicating that the conclusions are insensitive to the choice of photon‑number cutoff.
For both the simulated and experimental results shown in Fig.\ref{inequality_three_mode}, when $r\leq 0.53$, the three lines are very close, making it impossible to obtain any information about entanglement. When $r>0.53$, the red line falls below the lower bound of the RHS of the inequality, indicating a violation of the PPT inequality and thereby verifying the presence of entanglement among the three modes of the Gaussian state. 

In this work, we verify that intensity correlation moments, without the need for phase information, can be used to measure entanglement in two-mode and three-mode Gaussian states. This method can be extended to Gaussian states with more than three modes. For the classes studied here, entanglement can be assessed using intensity correlation moments without recourse to phase-sensitive measurements. Extending this approach to $N$-mode Gaussian states with $N>3$ is possible in principle, but requires access to additional invariants and higher-order moments. 
The results obtained for the two Gaussian states confirm that the entanglement characteristics of some Gaussian states can be verified without phase information. For non-Gaussian states, intensity correlation moments alone may not be sufficient to infer entanglement.
%

\section{\label{sec4}Conclusion}
In this work, we have derived the symplectic QUIs of the TMSV and TMGS from intensity correlation moments and used the PPT criterion to assess entanglement. Furthermore, we have obtained the symplectic eigenvalues of the TMSV, which further reveal the statistical characteristics of the state. Unlike homodyne detection~\cite{PhysRevLett.102.020502,yokoyama2013ultra} and entropy-based detection~\cite{islam2015measuring}, which require phase information or reference light that can be difficult to provide experimentally, the intensity correlation moment approach based on PNR detection~\cite{gondret2025quantifying,pevrina2020nonclassicality,pevrina2022nonclassicality} avoids these requirements.

We have experimentally demonstrated the entanglement characteristics of two-mode and three-mode Gaussian states without phase information. The experimental results show that both states violate the PPT criterion; however, due to the absence of phase information, we can only certify entanglement in the TMGS when $r>0.53$.

At the same time, the experimental results show that the smallest symplectic eigenvalue of the partially transposed covariance matrix $\tilde \lambda_-$ of the TMSV is below 1 for $r<0.96$, and it first decreases and then increases as $r$ grows. It may become larger than 1 when $r>0.96$ due to the multimode structure of the pump laser; this issue could be mitigated by increasing the number of detectors and using a purer pump laser.

\begin{acknowledgments}
Z.-S. He and Y. Zhao contributed equally to this paper. 
This work was supported by the National Key Research and Development Program of China (No.\,2025YFE0217700), the Quantum Science and Technology-National Science and Technology Major Project (No.\,2021ZD0301200), the National Natural Science Foundation of China (No.\,12474494), the CAS Project for Young Scientists in Basic Research (YSBR-131).
\end{acknowledgments}

\bibliography{ref}

\end{document}


\pagebreak

\onecolumngrid

\title{Supplementary Material for \emph{Experimental detection of entanglement in multimode Gaussian states from high-order intensity correlation moments}}

\author{Ze-Shan He}
\affiliation{Laboratory of Quantum Information, University of Science and Technology of China, Hefei 230026, China}
\affiliation{Hefei National Laboratory, University of Science and Technology of China, Hefei 230088, China}
\affiliation{Anhui Province Key Laboratory of Quantum Network, University of Science and Technology of China, Hefei 230026, China}
\affiliation{CAS Center for Excellence in Quantum Information and Quantum Physics, University of Science and Technology of China, Hefei 230026, China}
\author{Yukuan Zhao}
\affiliation{Laboratory of Quantum Information, University of Science and Technology of China, Hefei 230026, China}
\affiliation{Anhui Province Key Laboratory of Quantum Network, University of Science and Technology of China, Hefei 230026, China}
\affiliation{School of Physical Sciences, University of Science and Technology of China, Hefei 230026, China}
\author{Hao-Shu Tian}
\affiliation{Laboratory of Quantum Information, University of Science and Technology of China, Hefei 230026, China}
\affiliation{Hefei National Laboratory, University of Science and Technology of China, Hefei 230088, China}
\affiliation{Anhui Province Key Laboratory of Quantum Network, University of Science and Technology of China, Hefei 230026, China}
\affiliation{CAS Center for Excellence in Quantum Information and Quantum Physics, University of Science and Technology of China, Hefei 230026, China}
\author{Kai Sun}
\author{Xiao-Ye Xu}
\email{xuxiaoye@ustc.edu.cn}
\author{Chuan-Feng Li}
\email{cfli@ustc.edu.cn}
\author{Guang-Can Guo}
\affiliation{Laboratory of Quantum Information, University of Science and Technology of China, Hefei 230026, China}
\affiliation{Hefei National Laboratory, University of Science and Technology of China, Hefei 230088, China}
\affiliation{Anhui Province Key Laboratory of Quantum Network, University of Science and Technology of China, Hefei 230026, China}
\affiliation{CAS Center for Excellence in Quantum Information and Quantum Physics, University of Science and Technology of China, Hefei 230026, China}

\date{\today}

\setcounter{equation}{0}
\setcounter{figure}{0}
\setcounter{table}{0}
\setcounter{page}{1}
\makeatletter
\renewcommand{\theequation}{S\arabic{equation}}
\renewcommand{\thefigure}{S\arabic{figure}}
\renewcommand{\bibnumfmt}[1]{[#1]}
\renewcommand{\citenumfont}[1]{#1}

\maketitle

\newpage

\tableofcontents

\newpage

\section{Joint Spectral Intensity Distribution}
\begin{figure}[htbp]
    \centering
    \includegraphics[width=0.45\textwidth]{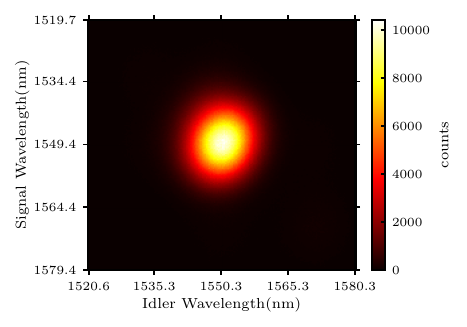}
    \caption{\textbf{Joint Spectral Intensity Distribution.}
    We measure the JSI distribution of the cpKTP crystal using a fiber-assisted single-photon spectrograph with two 4.5\,km DCF. The bandwidths of the signal and idler photons are \textcolor{red}{13.1}\,nm and \textcolor{red}{12.9}\,nm, respectively. The center wavelength is 1550.3\,nm. The deviation is due to a shift in the center wavelength of the pump laser. The purity and the Schmidt numbers of the JSI are 0.9945 and 1.0055, respectively.}
    \label{jsi}
\end{figure}

We generate spectrally pure biphotons using a customized quasi-phase-matched KTiOPO4 (CPKTP) crystal, which produces a Gaussian phase-matching function via a customized poling configuration\,\cite{PhysRevA.93.013801,PhysRevApplied.12.034059}. We use a time-domain fiber-assisted single-photon spectrograph\,\cite{avenhaus2009fiber} to measure the joint spectral intensity\,(JSI) distribution of the CPKTP crystal with two 4.5\,km dispersion-compensating fibers\,(DCF). These fibers map the spectrum from the frequency domain to the time domain, giving a spectral resolution of 0.19\,nm. The purity of the JSI reaches 0.9945, and the Schmidt number reaches 1.0055, indicating that there is nearly a single spectral mode, leaving little room for further improvement. The measured JSI of our CPKTP crystal is shown in Fig.\,(\ref{jsi}).

\section{Pseudo-Photon-Number-Resolving Detector}
\begin{figure}[htbp]
    \centering
    \includegraphics[width=0.45\textwidth]{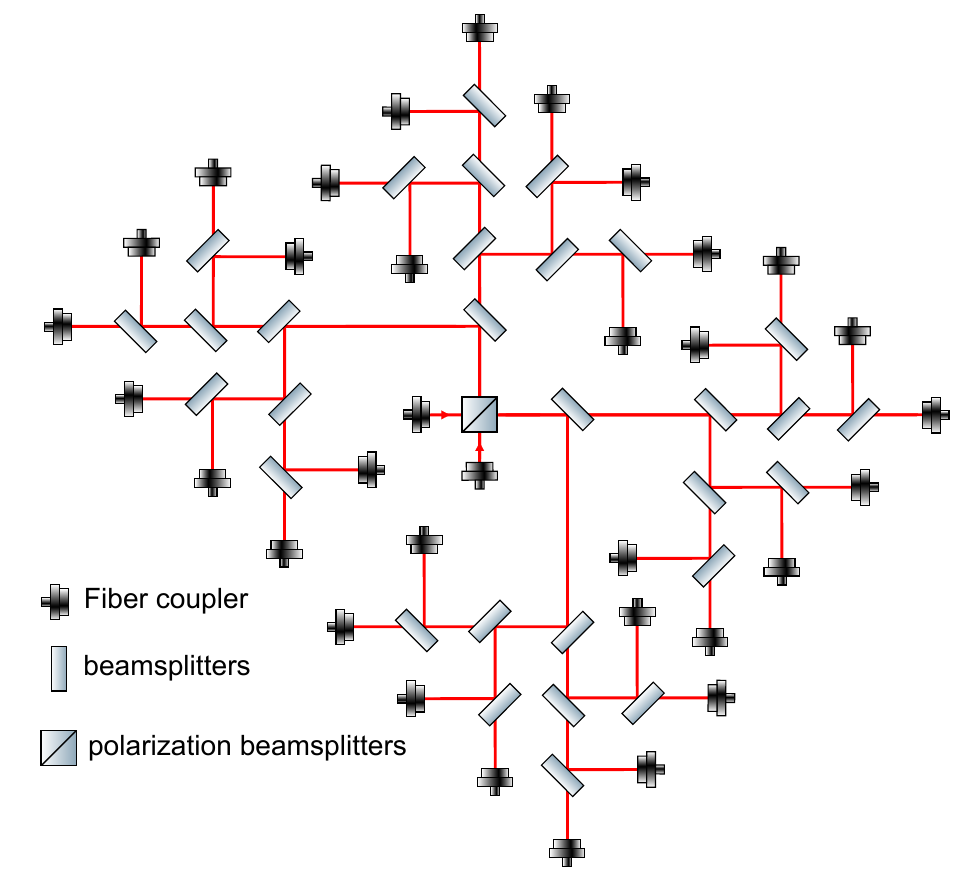}
    \caption{\textbf{32-channel photon-number-resolving detector.}
    The setup has two functionally equivalent input ports and 32 output ports. The input signal is divided into 32 paths by a network of beam splitters and subsequently detected by the PNR detector.}
    \label{pnr}
\end{figure}

In this work, the symplectic quantum universal invariants\,(QUIs) of the TMGS require intensity correlation moments up to sixth order, specifically $\langle W_1^2 W_2^2 W_3^2 \rangle$, which can be measured by a pseudo-photon-number-resolving\,(PNR) detector. Here, we build a 32-channel pseudo-PNR detector to access such high-order intensity correlation moments. The experimental setup is shown in Fig.(\ref{pnr}). All fiber couplers are connected to a superconducting nanowire single-photon detector\,(Scontel, 16COPRS-CCR-TW-90) via single-mode fiber. The total detection efficiency of the pseudo-PNR is $82\%$, which includes an $8\%$ detection loss, an $8\%$ coupling loss, and a $2\%$ fiber transmission loss.

\section{Comparison of the inequality for different photon‑number cutoffs in TMGS}

In this section, we present experimental and simulated results for photon‑number cutoffs of $8,11$ and $14$. The trends across all results are consistent, indicating that the conclusions in the main text are insensitive to the choice of photon‑number cutoff. The deviation between the experimental and simulated data is attributed to detection losses and imperfections in the pseudo‑PNR detector. 

\begin{figure}[htbp]
    \centering
    \includegraphics[width=1.0\textwidth]{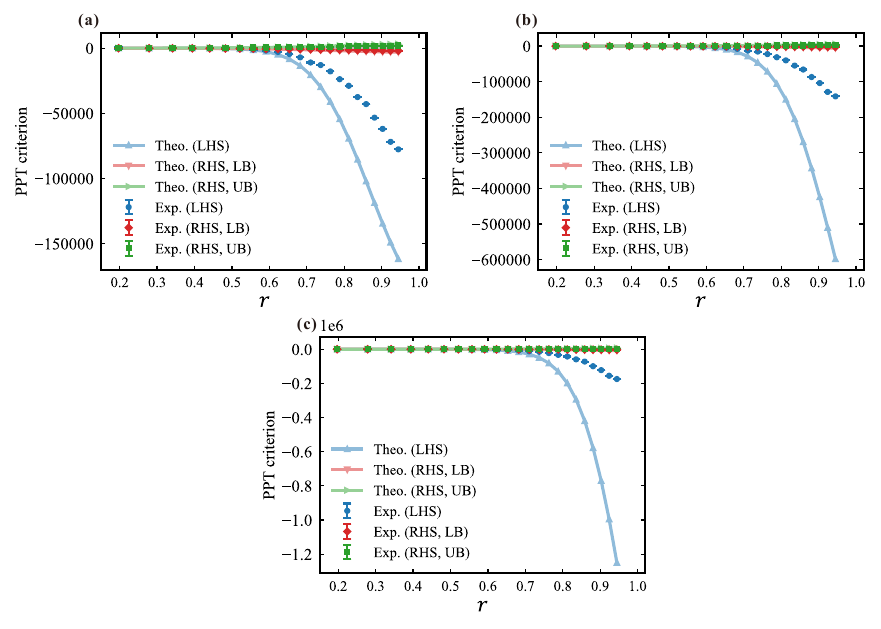}
    \caption{\textbf{Different cutoff of the inequality in TMGS.}
    Experimental and simulated results for the inequality as a function of $r$ at $g^2(0)=1.8$, for different photon-number cutoff of (a) 8, (b) 11, and (c) 14. The light blue solid line with up-triangle markers represents the simulated value of the left-hand side\,(LHS) of the inequality. The light green and light red solid lines with right- and down-triangle markers represent the upper bounds\,(UB) and lower bounds\,(LB) of the right-hand side\,(RHS), respectively. Experimental data are shown as corresponding dots\,(see legend).}
    \label{compare}
\end{figure}
\section{Conventions and Basic Objects}

We consider three bosonic modes described by annihilation/creation operators $\hat a_j,\hat a_j^\dagger$ that satisfy $[\hat a_j,\hat a_k^\dagger]=\delta_{jk}$. We adopt the quadrature convention
\begin{equation}
  \hat x_j = \frac{\hat a_j+\hat a_j^\dagger}{\sqrt{2}},\qquad
  \hat p_j = \frac{\hat a_j-\hat a_j^\dagger}{\sqrt{2}\,i},\qquad
  [\hat x_j,\hat p_k]= i\,\delta_{jk},
\end{equation}
so that the vacuum covariance is $\sigma_{\mathrm{vac}}=\frac{1}{2}\,\mathbb{I}_{2}$ per mode.

The \emph{normal-ordered characteristic function} (normal CF) for three modes is
\begin{equation}\label{eq:CN-def}
  C_N(\bm\beta) \equiv \Tr\!\left[\hat\rho\,e^{\beta_1\hat a_1^\dagger}e^{-\beta_1^*\hat a_1}\,
                                 e^{\beta_2\hat a_2^\dagger}e^{-\beta_2^*\hat a_2}\,
                                 e^{\beta_3\hat a_3^\dagger}e^{-\beta_3^*\hat a_3}\right],\qquad
  \bm\beta=(\beta_1,\beta_2,\beta_3)\in\mathbb C^3.
\end{equation}
Expanding the exponentials shows that $C_N$ is the generating function of the normally ordered moments\,\cite{perina1991quantum1}:
\begin{equation}\label{eq:gen-moment}
  C_N(\bm\beta)= \sum_{\bm m,\bm n\ge 0}
     \prod_{j=1}^3 \frac{\beta_j^{m_j} (-\beta_j^*)^{n_j}}{m_j!\,n_j!}\,
     \langle (\hat a_1^\dagger)^{m_1}\hat a_1^{n_1}\,
             (\hat a_2^\dagger)^{m_2}\hat a_2^{n_2}\,
             (\hat a_3^\dagger)^{m_3}\hat a_3^{n_3}\rangle.
\end{equation}
Therefore,
\begin{equation}\label{eq:moment-deriv-master}
  \boxed{\ \langle (\hat a^\dagger)^{\bm m}\hat a^{\bm n}\rangle
  = (-1)^{n_1+n_2+n_3}\!
     \left.\prod_{j=1}^3 \partial_{\beta_j}^{m_j}\,\partial_{\beta_j^*}^{n_j}\,C_N(\bm\beta)\right|_{\bm\beta=\bm 0} }
\end{equation}
(where the multi-index notation is obvious).

For a (zero-mean) TMGS, $C_N$ is a quadratic exponential in $\beta_j,\beta_j^*$:
\begin{align}
  C_N(\bm\beta) = \exp Q(\bm\beta),\qquad
  Q(\bm\beta) = &-\sum_{j=1}^3 B_j \abs{\beta_j}^2
  -\sum_{j=1}^3\left(\frac{C_j}{2}\beta_j^{*2}+\frac{C_j^*}{2}\beta_j^{2}\right)\nonumber\\
  &+\sum_{1\le j<k\le 3}\!\!\left(D_{jk}\,\beta_j^*\beta_k^*+D_{jk}^*\,\beta_j\beta_k
                    +\bar D_{jk}\,\beta_j\beta_k^*+\bar D_{jk}^*\,\beta_j^*\beta_k \right).
  \label{eq:Q-3m}
\end{align}
Here $B_j\in\mathbb R$ and $C_j,D_{jk},\bar D_{jk}\in\mathbb C$.
The absence of terms beyond quadratic order reflects the fact that a Gaussian state is completely characterized by its first and second moments.

\section{Determinant of the TMSV covariance matrix}
The determinant of the covariance matrix of a TMSV state can be expressed in terms of intensity correlation moments~\cite{barasinski2023quantification}:
\begin{equation}
    \begin{aligned}
    \det \sigma = &\; 1 + 4(\langle W_1 \rangle + \langle W_2 \rangle) + 12(\langle W_1 \rangle + \langle W_2 \rangle)^2 - 4\langle W_1^2 \rangle(1 + 6\langle W_2 \rangle + 24\langle W_2 \rangle^2) \\
                  &- 4\langle W_2^2 \rangle(1 + 6\langle W_1 \rangle + 24\langle W_1 \rangle^2) + 8\langle W_1^2 W_2 \rangle(1 + 6\langle W_2 \rangle) + 8\langle W_1 W_2^2 \rangle(1 + 6\langle W_1 \rangle) \\
                  &- 8\langle W_1 W_2 \rangle(1 + 6\langle W_1 \rangle + 6\langle W_2 \rangle + 48\langle W_1 \rangle\langle W_2 \rangle) + 96\langle W_1 \rangle\langle W_2 \rangle(\langle W_1 \rangle + \langle W_2 \rangle + 5\langle W_1 \rangle\langle W_2 \rangle) \\
                  &+ 24\langle W_1^2 \rangle\langle W_2^2 \rangle - 8\langle W_1^2 W_2^2 \rangle + 48\langle W_1 W_2 \rangle^2.
    \end{aligned}
    \label{eq:det_sigma}
\end{equation}, where $W_j = \hat a_j^\dagger \hat a_j$ is the normally ordered intensity operator.
\section{Derivation of the PPT Criterion for TMGS}
\subsection{Covariance Matrix of General TMGS}

We consider a general TMGS on modes $j=1,2,3$. Its covariance matrix in the standard ordering
$\hat{\boldsymbol{\xi}}^{\mathsf T} = (\hat x_1,\hat p_1,\hat x_2,\hat p_2,\hat x_3,\hat p_3)$
has the usual block form
\begin{equation}
    \begin{aligned}
    \sigma =
    \begin{bmatrix}
    \sigma_1 & \gamma_{12} & \gamma_{13} \\
    \gamma_{12}^T & \sigma_2 & \gamma_{23} \\
    \gamma_{13}^T & \gamma_{23}^T & \sigma_3 
    \end{bmatrix}
    \end{aligned}
    \label{convariance_matrix_for_threeSM}
\end{equation}
where the $\sigma_j$ are the $2\times 2$ local covariance matrices and
$\gamma_{jk}$ are the $2\times 2$ inter-mode correlation blocks. The local covariance matrices and correlation blocks are parameterized by

\begin{equation}
    \begin{aligned}
    \sigma_j &=
    \begin{bmatrix}
    1 + 2B_j + 2\Re C_j & 2\Im C_j \\
    2\Im C_j & 1 + 2B_j - 2\Re C_j
    \end{bmatrix}, \\
    \gamma_{jk} &=
    \begin{bmatrix}
    2\Re(D_{jk}-\bar D_{jk}) & 2\Im(D_{jk}-\bar D_{jk}) \\
    2\Im(D_{jk}+\bar D_{jk}) & -2\Re(D_{jk}+\bar D_{jk})
    \end{bmatrix}.
    \end{aligned}
    \label{block_matrix}
\end{equation}
\title{Detailed Derivation of \(\tilde{\Delta}_1^3\) for General TMGS}

The normal-ordered parameters $B_j \in \mathbb{R}$ and
$C_j, D_{jk}, \bar D_{jk}\,(j\neq k)\in \mathbb{C}$ are defined by
\begin{equation}
  B_j = \langle  \hat a_j^\dagger  \hat a_j\rangle,\qquad
  C_j = \langle ( \hat a_j)^2\rangle,\qquad
  D_{jk} = \langle  \hat a_j \hat a_k\rangle,\qquad
  \bar D_{jk} = - \langle  \hat a_j^\dagger  \hat a_k\rangle,
\end{equation}
for $j,k = 1,2,3$, $j<k$, where the normally ordered intensity operators $W_j = \hat a_j^\dagger \hat a_j$ obey the relations
\begin{equation}
\begin{aligned}
  \langle W_j \rangle &= B_j,\\
  \langle W_j^2 \rangle &= 2 B_j^2 + |C_j|^2,\\
  \langle W_j W_k \rangle &= B_j B_k + |D_{jk}|^2 + |\bar D_{jk}|^2,
  \quad j\neq k.
\label{seconde_moments}
\end{aligned}
\end{equation}

It is not sufficient to derive symplectic QUIs of TMGS using only lower-order moments; higher-order moments also provide valuable information. The third- and fourth-order intensity correlation moments yield additional information about the parameters and can be written in terms of
\begin{equation}
\begin{aligned}
-4\Re(C_1 \bar{D}_{12}D_{12}^\ast )=&-4\langle W_1\rangle(\langle W_1 W_2\rangle-\langle W_1\rangle\langle W_2\rangle)\\ &+\langle W_1^2 W_2\rangle-\langle W_1^2 \rangle\langle W_2\rangle\\
-4\Re(C_2^\ast \bar{D}_{12}D_{12})=&-4\langle W_2\rangle(\langle W_1 W_2\rangle-\langle W_1\rangle\langle W_2\rangle)\\ &+\langle W_1 W_2^2\rangle-\langle W_1 \rangle\langle W_2^2\rangle\\
4(2|D_{12}|^2|\bar{D}_{12}|^2+\Re(C_1 C_2^\ast & \bar{D}_{12}^2)+\Re(C_1 C_2 (D_{12}^\ast )^2))=\\
 \langle W_1^2 W_2^2\rangle-4\langle W_1 W_2\rangle ^2 &-8\langle W_1 \rangle\langle W_2\rangle \langle W_1 W_2\rangle +8\langle W_1\rangle ^2 \langle W_2\rangle ^2\\
-2[\langle W_1\rangle ^2(\langle W_2^2\rangle -2\langle W_2\rangle ^2)&+\langle W_2\rangle ^2(\langle W_1^2\rangle -2\langle W_1\rangle ^2)]\\
 -(\langle W_2^2\rangle -2\langle W_2\rangle ^2)&(\langle W_1^2\rangle -2\langle W_1\rangle ^2)\\
+16\langle  W_2\rangle \Re(C_1 \bar{D}_{12}D_{12}^\ast )&+16\langle  W_1\rangle \Re(C_2^\ast \bar{D}_{12}D_{12}).
\label{hightmoment}
\end{aligned}
\end{equation}

\subsection{Quantum universal invariants of TMGS}\label{sub:B}
In the main text, we introduced the PPT criterion based on the partially transposed covariance matrix. In Sec\,\ref{sub:B} and Sec\,\ref{sub:C} we show how to derive it from the symplectic QUIs $\Delta_k^N$.  

For an $N$-mode Gaussian state, the invariant $\Delta_k^N$ is defined as the principal minors of order $2k$ of the matrix $\Omega \sigma$, where
$\Omega = \bigoplus_{k=1}^N \begin{pmatrix} 0 & 1 \\ -1 & 0 \end{pmatrix}$. The general expression for the invariant is 
\begin{equation}
  \Delta_k^N := M_{2k}(\Omega \sigma),
\end{equation}
and they determine the symplectic eigenvalues of $\sigma$. 

For a TMGS, the first symplectic QUI can be expressed as
\begin{equation}
\begin{aligned}
  \Delta_1^{3} &=\det \sigma_1 + \det \sigma_2 + \det \sigma_3 + 2 \left( \det \gamma_{12} + \det \gamma_{13} + \det \gamma_{23} \right)\\
  &= \Delta_{1,w}^{3} - \Delta_{1,\beta}^{3},
\label{first_symplectic_invariants}
\end{aligned}
\end{equation}
where $\Delta_{1,w}^{3}$ equals the sum of the determinants of three diagonal block matrices\,\cite{barasinski2023quantification}, and $\Delta_{1,\beta}^{3}$ can be expressed by
\begin{equation}
  \Delta_{1,\beta}^{3}
  = 8\Bigl[
      \bigl(|D_{12}|^2 - |\bar D_{12}|^2\bigr)
    + \bigl(|D_{13}|^2 - |\bar D_{13}|^2\bigr)
    + \bigl(|D_{23}|^2 - |\bar D_{23}|^2\bigr)
    \Bigr],
  \label{eq:Delta3-1r-explicit}
\end{equation} 
where we define $r_{ij}=|D_{ij}|^2 - |\bar D_{ij}|^2$, those parameters can be derived from the second- and fourth-order intensity correlations $g^{(2)}_{ij}$ and $g^{(4)}_{ij}$ in a TMGS with $i,j \in [1,2,3]$\,\cite{gondret2025quantifying}.

The second symplectic QUI is written as
\begin{equation}
  \Delta_2^{3}
   = \Delta_{2,w}^{3} - \Delta_{2,r}^{3},
\end{equation}
where $\Delta_{2,w}^{3}$ equals the sum of the second symplectic QUI of the TMGS. We define
\begin{align}
F(1,2,3)
&=
16(B_1+B_1^2-|C_1|^2)\Bigl(|D_{23}|^2-|\bar D_{23}|^2\Bigr)
+16\,\Re\!\Bigl(D_{12}D^*_{13}\bar D_{23}\Bigr)
\nonumber\\
&\quad
+32(B_1+B_2-B_3)\,\Re\!\Bigl(\bar D_{12}D^*_{23}D_{13}\Bigr)
-16(2B_1+1)\,\Re\!\Bigl(\bar D_{12}\bar D^*_{13}\bar D_{23}\Bigr)
\nonumber\\
&\quad
-16\Bigl(|D_{12}|^2-|\bar D_{12}|^2\Bigr)\Bigl(|D_{13}|^2-|\bar D_{13}|^2\Bigr)
\nonumber\\
&\quad
+32\Bigl(
\Re\!\bigl(C_1D^*_{12}D^*_{13}D_{23}\bigr)
+\Re\!\bigl(C_1\bar D_{12}\bar D_{13}D^*_{23}\bigr)
\Bigr)
-64\,\Re\!\Bigl(C_2\bar D_{13}\bar D^*_{12}D^*_{23}\Bigr),
\label{eq:F123}
\end{align}
Thus the residual terms $\Delta_{2,r}^{3}$ can be expressed by
\begin{equation}
\begin{split}
  \Delta_{2,r}^{3}
  &= \Bigl\{ 8\bigl(|D_{12}|^2\bigr) + \mathrm{addt2} \Bigr\} \quad + \Bigl\{
      F(1,2,3) + \mathrm{addt3}
     \Bigr\}.
\end{split}
\label{eq:Delta3-2r-def}
\end{equation}

The symbol $\mathrm{addt2}$ denotes all the terms obtained by replacing the index pair $\{1,2\}$ with each of the pairs $\{2,1\},\{1,3\},$ $\{3,1\},\{2,3\}$, and $\{3,2\}$. Similarly, $\mathrm{addt3}$ denotes the terms obtained by all nontrivial permutations of the triplet $\{1,2,3\}$, i.e., $\{1,3,2\},\{2,1,3\},\{3,1,2\},\{2,3,1\}$, and $\{3,2,1\}$. These contributions together provide the exact expression for the residual terms of $\Delta_2^3$ for a general TMGS. 

The third symplectic QUI is simply $\Delta_3^3=\det\sigma$. Note that, for a TMGS, the third symplectic QUI $\Delta_3^{3}$ can be fully expressed in terms of intensity correlation moments, and there is no residual term for $\Delta_3^{3}$\,\cite{sudak2025efficient}.

\subsection{Necessary Condition Separability from  Partial Transposition State}\label{sub:C}

For any separable TMGS, the partially transposed covariance matrix must satisfy the uncertainty relation, which gives the necessary condition
\begin{equation}
  \tilde\Delta_3^{3} - \tilde\Delta_2^{3} + \tilde\Delta_1^{3} - 1 \;\ge\; 0.
  \label{eq:RS-3mode_pt}
\end{equation}
where $\tilde\Delta_k^{N}$ denotes the symplectic QUI of the partially transposed covariance matrix, which can be obtained from the symplectic QUI of the TMGS. Violation of this inequality certifies entanglement across the corresponding bipartition.

We now consider partial transposition applied to mode 1 to obtain~\eqref{eq:RS-3mode_pt}. Partial transposition with respect to other modes produces analogous results. Partial transposition with respect to mode 1 flips the sign of the momentum quadrature \(p_1\) corresponding to the transformation $\tilde{\sigma} = \Lambda \sigma \Lambda$, where $\Lambda = \text{diag}(1, -1, 1, 1, 1, 1)$.

\[
\tilde{\sigma}_j = \sigma_j \quad \text{for} \quad j = 2, 3 \quad \text{(unchanged)}, \quad \det {\tilde\sigma}_1 = \det \sigma_1.
\]
\[
\det \gamma_{jk} = -4 \left( |D_{jk}|^2 - | \bar{D}_{jk}|^2 \right).
\]
Define \(r_{jk} = |D_{jk}|^2 - |\bar{D}_{jk}|^2\) for convenience. Then,
\[
\det \gamma_{jk} = -4 r_{jk}.
\]
For \(\gamma_{12}\) and \(\gamma_{13}\), the bottom row signs flip, yielding \(\det \tilde{\gamma}_{12} = -\det \gamma_{12}\) and \(\det \tilde{\gamma}_{13} = -\det \gamma_{13}\). As for \(\gamma_{23}\), it does not involvement of beam 1, so \(\det \tilde{\gamma}_{23} = \det \gamma_{23}\). Combining with ~\eqref{first_symplectic_invariants}, we can obtain


\[
\tilde{\Delta}_1^3 - \Delta_1^3 = 2 \left( - \det \gamma_{12} - \det \gamma_{13} + \det \gamma_{23} \right)-2 \left(  \det \gamma_{12} + \det \gamma_{13} + \det \gamma_{23} \right)  = - 4 \left( \det \gamma_{12} + \det \gamma_{13} \right)
\]

\[
-4 \left( \det \gamma_{12} + \det \gamma_{13} \right) = 16 \left( r_{12} + r_{13} \right).
\]
Thus,
\[
\tilde{\Delta}_1^3 = \Delta_1^3 + 16 \left( r_{12} + r_{13} \right).
\]
Using~\eqref{eq:Delta3-1r-explicit}, we obtain
\[
16 \left( r_{12} + r_{13} \right) = 2 \times 8 \left( r_{12} + r_{13} \right) = 2 \left( \Delta_{1,\beta}^3 - 8 \times r_{23} \right),
\]
and the final result can be expressed as
\begin{equation}
\tilde{\Delta}_1^3 = \Delta_1^3 + 2 \left( \Delta_{1,\beta}^3 - 8 \times r_{23} \right).
\end{equation}

The second symplectic QUI for a TMGS is defined as
\begin{equation}
\Delta^3_2 = M_4(\Omega \sigma),
\end{equation}
where $M_4$ denotes the sum of all principal minors of order $4$ of $\Omega \sigma$.
For a general TMGS, this quantity admits the decomposition
\begin{equation}
\Delta^3_2 = \Delta^3_{2,w} - \Delta^3_{2,r},
\end{equation}
where $\Delta^3_{2,w}$ depends only on intensity correlation moments and $\Delta^3_{2,r}$ contains all phase--sensitive contributions. 

Partial transposition with respect to mode $1$ corresponds, in phase space, to time reversal on that mode:$x_1 \mapsto x_1, p_1 \mapsto -p_1$. This is implemented by the diagonal matrix
\begin{equation}
\Lambda_1 = \mathrm{diag}(1,-1,\,1,1,\,1,1),
\end{equation}
so that the partial transposition covariance matrix is
\begin{equation}
\tilde \sigma = \Lambda_1 \sigma \Lambda_1 .
\end{equation}
Now, let $T = \mathrm{diag}(1,-1)$, thus the blockwise action can expressed as
\begin{equation}
    \begin{aligned}
    \tilde{\sigma}_1 &= T \sigma_1 T, \\
    \tilde{\gamma}_{1k} &= T \gamma_{1k}, \qquad
    \tilde{\gamma}_{k1} = \gamma_{k1} T, \quad (k=2,3), \\
    \tilde{\sigma}_2 &= \sigma_2, \quad
    \tilde{\sigma}_3 = \sigma_3, \quad
    \tilde{\gamma}_{23} = \gamma_{23}.
    \end{aligned}
    \label{blockwise_action}
\end{equation}

Since $\Delta^3_{2,w}$ depends only on intensity correlation moments, it is invariant under
partial transposition:
\begin{equation}
\tilde{\Delta}^3_{2,w} = \Delta^3_{2,w},
\end{equation}
therefore, the second symplectic QUI after partial transposition can be expressed by
\begin{equation}
\tilde{\Delta}^3_2 = \Delta^3_{2,w} - \tilde{\Delta}^3_{2,r},
\end{equation}
where $\tilde{\Delta}^3_{2,r}$ is obtained by substituting the transformed parameters into ~\eqref{eq:Delta3-2r-def}.

\paragraph{Local squeezing parameter $C_1\to C_1^*$.}
\[
\tilde\sigma_1=T\sigma_1T
=
\begin{pmatrix}1&0\\0&-1\end{pmatrix}
\begin{pmatrix}
1+2B_1+2\Re C_1 & 2\Im C_1\\
2\Im C_1 & 1+2B_1-2\Re C_1
\end{pmatrix}
\begin{pmatrix}1&0\\0&-1\end{pmatrix}.
\]

Because left multiplication by $T$ flips the sign of the second row, and right multiplication
by $T$ flips the sign of the second column, the diagonal entries are unchanged, while the off-diagonal
entries change sign:
\[
\tilde\sigma_1=
\begin{pmatrix}
1+2B_1+2\Re C_1 & -2\Im C_1\\
-2\Im C_1 & 1+2B_1-2\Re C_1
\end{pmatrix}.
\]

Comparison with the structural form of $\tilde\sigma_1$ yields
\[
\tilde B_1=B_1,\qquad \Re\tilde C_1=\Re C_1,\qquad \Im\tilde C_1=-\Im C_1,
\]
Hence, we can obtain
\begin{equation}
\tilde C_1=C_1^*.
\label{eq:C1_map}
\end{equation}

Moreover, $\tilde\sigma_2=\sigma_2$ and $\tilde\sigma_3=\sigma_3$ imply
$\tilde B_2=B_2$, $\tilde C_2=C_2$, $\tilde B_3=B_3$, $\tilde C_3=C_3$.

\paragraph{Correlation parameters $(D_{1k},\bar D_{1k})$ for $k=2,3$.}

Here, we fixed $k\in\{2,3\}$; thus, we can define the combinations
\begin{equation}
u:=D_{1k}-\bar D_{1k},\qquad v:=D_{1k}+\bar D_{1k}.
\label{eq:uv_def}
\end{equation}

Then the $\gamma_{ik}$ in \eqref{block_matrix} reads
\begin{equation}
\gamma_{1k}=
\begin{pmatrix}
2\Re u & 2\Im u\\
2\Im v & -2\Re v
\end{pmatrix}.
\label{eq:eps_uk_vk}
\end{equation}

Using~\eqref{blockwise_action}, the partially transposed correlation block is
\[
\tilde\gamma_{1k}=T\gamma_{1k}
=
\begin{pmatrix}1&0\\0&-1\end{pmatrix}
\begin{pmatrix}
2\Re u & 2\Im u\\
2\Im v & -2\Re v
\end{pmatrix}
=
\begin{pmatrix}
2\Re u & 2\Im u\\
-2\Im v & 2\Re v
\end{pmatrix}.
\]

On the other hand, $\tilde\gamma_{1k}$ covariance matrix can be parametrized in the same algebraic form as ~\eqref{eq:eps_uk_vk} in terms of the \emph{transformed} parameters
$\tilde D_{1k}$ and $\tilde{\bar D}_{1k}$. Define
\[
\tilde u:=\tilde D_{1k}-\tilde{\bar D}_{1k},\qquad
\tilde v:=\tilde D_{1k}+\tilde{\bar D}_{1k}.
\]

Then
\[
\tilde\gamma_{1k}=
\begin{pmatrix}
2\Re \tilde u & 2\Im \tilde u\\
2\Im \tilde v & -2\Re \tilde v
\end{pmatrix}.
\]

Comparing the resulting expression entry-by-entry with the explicit form of $T\varepsilon_{1k}$ given above yields
\begin{equation}
\tilde u=u,\qquad \tilde v=-v.
\label{eq:u_v_constraints}
\end{equation}

Solving~\eqref{eq:u_v_constraints} for $(\tilde D_{1k},\tilde{\bar D}_{1k})$ gives
\[
\tilde D_{1k}=\frac{\tilde u+\tilde v}{2}=\frac{u-v}{2}=-\bar D_{1k},\qquad
\tilde{\bar D}_{1k}=\frac{\tilde v-\tilde u}{2}=\frac{-v-u}{2}=-D_{1k}.
\]

We obtain the partial transpose of $D_{jk}$
\begin{equation}
\tilde D_{1k}=-\bar D_{1k},\qquad
\tilde{\bar D}_{1k}=-D_{1k}\qquad (k=2,3),
\label{eq:DbarD_map}
\end{equation}
Substituting~\eqref{eq:Delta3-2r-def},~\eqref{eq:C1_map} and ~\eqref{eq:DbarD_map} into the residual terms of the second symplectic QUIs, we find that the partially transposed residual is
\begin{equation}
\tilde{\Delta}^3_{2,r}
=
16\bigl(|\bar D_{12}|^2 + |\bar D_{13}|^2 + |D_{23}|^2\bigr)
+
\sum_{\pi\in S_3} \tilde F(\pi(1),\pi(2),\pi(3)),
\end{equation}
where $\tilde F$ can be expressed by ~\eqref{eq:F_123}, with pairs not involving mode 1 unchanged, and $\pi$ denotes a permutation.
\begin{align}
\tilde F(1,2,3)=\;&16(B_1+B_1^2-|C_1|^2)\Big(|D_{23}|^2-|\bar D_{23}|^2\Big)\nonumber\\
&+16\,\Re\!\Big(\bar D_{12}\, \bar D_{13}^*\, \bar D_{23}\Big)
+32(B_1+B_2-B_3)\Re\!\Big(D_{12}\,\bar D_{13}\,D_{23}^*\Big)\nonumber\\
&-16(2B_1+1)\Re\!\Big(D_{12}\,D_{13}^*\,\bar D_{23}\Big)
-16\Big(|D_{12}|^2-|\bar D_{12}|^2\Big)\Big(|D_{13}|^2-|\bar D_{13}|^2\Big)\nonumber\\
&+32\Big(\Re\!\big(C_1^*\,D_{12}D_{13}D_{23}^*\big)+\Re\!\big(C_1^*\,\bar D_{12}^*\bar D_{13}^*D_{23}\big)\Big)
-64\,\Re\!\Big(C_2\,D_{13}D_{12}^*D_{23}^*\Big).
\label{eq:F_123}
\end{align}

The general expression for the partially transposed invariant is given by
\begin{equation}
\tilde{\Delta}^3_2
=
\Delta^3_{2,w}
-
\left[
16\bigl(|\bar D_{12}|^2 + |\bar D_{13}|^2 + |D_{23}|^2\bigr)
+
\sum_{\pi\in S_3} \tilde F(\pi(1),\pi(2),\pi(3))
\right].
\end{equation}

This result holds for any general TMGS and follows directly from the definition of $\Delta^3_2$ and the phase space action of partial transposition. Using the partially transposed symplectic QUIs, we can express ~\eqref{eq:RS-3mode_pt} by substituting the above results, which yields

$$\Delta^{3}_{3}-\Delta^3_{2,w}
+
\left[
16\bigl(|\bar D_{12}|^2 + |\bar D_{13}|^2 + |D_{23}|^2\bigr)
+
\sum_{\pi\in S_3} \tilde F(\pi(1),\pi(2),\pi(3))
\right]+\left[\Delta_1^3 + 2 \left( \Delta_{1,\beta}^3 - 8 \times r_{23} \right)\right]\ge 1,$$
placing the terms that can be expressed in terms of intensity correlation moments on the left-hand side and the rest on the right-hand side yields the inequality
\begin{equation}
\Delta^{3}_{3}-\Delta^3_{2,w}
+\Delta_1^3 + 2 \Delta_{1,\beta}^3 - 16r_{23} -1 + 16\bigl(|\bar D_{12}|^2 + |\bar D_{13}|^2 + |D_{23}|^2\bigr)\ge
-\sum_{\pi\in S_3} \tilde F(\pi(1),\pi(2),\pi(3)).
\label{eq:ppt criterion}
\end{equation}
with \(r_{jk} = |D_{jk}|^2 - |\bar{D}_{jk}|^2\). The $\tilde{F}$ terms are defined above.

\subsection{Upper and Lower Bound of the Right-hand Side of the Inequality}
Due to the loss of phase information, the terms in ~\eqref{eq:F_123} cannot be expressed in terms of intensity correlation moments. We therefore derive the upper and lower bounds using $|D_{jk}|^{2},|\bar D_{jk}|^{2}$. Consider the bounds of $-\tilde F(1,2,3)$.

It is convenient to define the non-negative, covariance-like quantities as
\begin{equation}
  \Sigma_{jk} \equiv \langle W_j W_k\rangle - \langle W_j\rangle\langle W_k\rangle
  = |D_{jk}|^{2} + |\bar D_{jk}|^{2} \ge 0.
  \label{eq:Sigma-def}
\end{equation}

From ~\eqref{eq:Sigma-def} we immediately obtain the bounds
\begin{equation}
  |D_{jk}|^{2} \le \Sigma_{jk},\qquad
  |\bar D_{jk}|^{2} \le \Sigma_{jk},\qquad
  |D_{jk}|,\ |\bar D_{jk}|
  \le \sqrt{\Sigma_{jk}}.
  \label{eq:D-bounds}
\end{equation}
Moreover,~\eqref{eq:Sigma-def} implies the key inequality\,(Eq.~(13) in\,\cite{sudak2025efficient})
\begin{equation}
  \bigl||D_{jk}|^{2} - |\bar D_{jk}|^{2}\bigr|
  \le |D_{jk}|^{2} + |\bar D_{jk}|^{2}
  = \Sigma_{jk}.
  \label{eq:D-diff-bound}
\end{equation}



Next, using Eq.~(4) of\,\cite{sudak2025efficient}, the single-mode RS uncertainty relation $\det\sigma_j \ge 1$ for each beam $j$ yields,
\begin{equation}
  \det\sigma_j
  = 1 + 4B_j + 4B_j^{2} - 4|C_j|^{2} \ge 1
  \quad\Rightarrow\quad
  B_j^{2} + B_j - |C_j|^{2} \ge 0.
\end{equation}

Using the intensity identity $|C_j|^{2}=\langle W_j^{2}\rangle-2B_j^{2}$ from ~\eqref{seconde_moments}, we can rewrite this RS combination exactly as
\begin{equation}
  B_j + B_j^{2} - |C_j|^{2}
  = \bigl(B_j+B_j^{2}\bigr) - \bigl(\langle W_j^{2}\rangle-2B_j^{2}\bigr)
  = 3B_j^{2} + B_j - \langle W_j^{2}\rangle \;\ge\; 0.
  \label{eq:RS-intensity-exact}
\end{equation}

This is the key point: the RS constraint is now expressed entirely in
terms of the intensity moments $B_j=\langle W_j\rangle$ and $\langle W_j^2\rangle$.

Finally, from ~\eqref{seconde_moments} we also have
\begin{equation}
  |C_j|^{2} = \langle W_j^{2}\rangle - 2B_j^{2}
  = \langle W_j^{2}\rangle - 2\langle W_j\rangle^{2},
\end{equation}
so that
\begin{equation}
  |C_j| = \sqrt{\langle W_j^{2}\rangle - 2\langle W_j\rangle^{2}}.
  \label{eq:C-exact-intensity}
\end{equation}

In what follows, we also repeatedly use the elementary estimates
\begin{equation}
  |\mathrm{Re}\,z|\le |z|,\qquad
  |z_1 z_2 \cdots z_m| = |z_1|\cdots|z_m|,
  \label{eq:basic-CS}
\end{equation}
valid for all complex numbers $z,z_1,\dots,z_m$.


We now derived the upper and lower bounds for each term in $\tilde F_{123}$ of ~\eqref{eq:F_123} separately, and
express the result entirely in terms of $B_j$, $\langle W_j^2\rangle$ and
$\Sigma_{jk}$.

\paragraph{Term with $B_1 + B_1^{2} - |C_1|^{2}$.}

The first term of $R_{123}$ reads as
\begin{equation}
  T_1
  = 16\bigl(B_1 + B_1^{2} - |C_1|^{2}\bigr)\bigl(|D_{23}|^{2} - |\bar D_{23}|^{2}\bigr).
\end{equation}

Using ~\eqref{eq:D-diff-bound}, we obtain the tighter intensity-only
bound
\begin{equation}
  |T_1|
  \le 16\bigl|B_1^{2}+B_1-|C_1|^{2}\bigr|\,\Sigma_{23}.
  \label{eq:T1-tight-intensity}
\end{equation}

\paragraph{Purely cubic term $D_{12}D_{13}^{*}\bar D_{23}$.}

The second term reads as
\begin{equation}
  T_2 = 16\,\mathrm{Re}\{\bar D_{12}\bar D_{13}^{*}\bar D_{23}\}.
\end{equation}
Using ~\eqref{eq:basic-CS} and ~\eqref{eq:D-bounds}, we can obtain
\begin{align}
  |T_2|
  &\le 16\,|\bar D_{12}|\,|\bar D_{13}|\,|\bar D_{23}|
  \nonumber\\
  &\le 16\,
      \sqrt{\Sigma_{12}}\,
      \sqrt{\Sigma_{13}}\,
      \sqrt{\Sigma_{23}}
  = 16\bigl(\Sigma_{12}\Sigma_{13}\Sigma_{23}\bigr)^{1/2}.
  \label{eq:T2-bound}
\end{align}

\paragraph{Term with $B_1 + B_2 - B_3$.}

The third term is
\begin{equation}
  T_3
  = 32(B_1 + B_2 - B_3)\,\mathrm{Re}\{D_{12}D_{23}^{*}\bar D_{13}\}.
\end{equation}

Then, using ~\eqref{eq:basic-CS} and ~\eqref{eq:D-bounds},
\begin{align}
  |T_3|
  &\le 32|B_1 + B_2 - B_3|\,
      |D_{12}|\,|D_{23}|\,|\bar D_{13}|
  \nonumber\\
  &\le 32|B_1 + B_2 - B_3|\,
      \bigl(\Sigma_{12}\Sigma_{13}\Sigma_{23}\bigr)^{1/2}.
  \label{eq:T3-bound}
\end{align}

\paragraph{Term with $(2B_1+1)\,\bar D_{12}\bar D_{13}^{*}\bar D_{23}$.}

The fourth term reads
\begin{equation}
  T_4 = -16(2B_1+1)\,\mathrm{Re}\{D_{12}D_{13}^{*}\bar D_{23}\}.
\end{equation}
Since $2B_1+1 > 0$, we obtain
\begin{align}
  |T_4|
  &\le 16(2B_1+1)\,|D_{12}|\,|D_{13}|\,|\bar D_{23}|
  \nonumber\\
  &\le 16(2B_1+1)\,
      \bigl(\Sigma_{12}\Sigma_{13}\Sigma_{23}\bigr)^{1/2}.
  \label{eq:T4-bound}
\end{align}

\paragraph{Quadratic product of $(|D|^{2}-|\bar D|^{2})$.}

The fifth term is
\begin{equation}
  T_5
  = -16\bigl(|D_{12}|^{2} - |\bar D_{12}|^{2}\bigr)
        \bigl(|D_{13}|^{2} - |\bar D_{13}|^{2}\bigr).
\end{equation}
Using ~\eqref{eq:D-diff-bound},
\begin{align}
  |T_5|
  &\le 16\,
      \bigl||D_{12}|^{2} - |\bar D_{12}|^{2}\bigr|
      \bigl||D_{13}|^{2} - |\bar D_{13}|^{2}\bigr|
  \nonumber\\
  &\le 16\,\Sigma_{12}\,\Sigma_{13}.
  \label{eq:T5-bound}
\end{align}

\paragraph{Terms involving $C_1$ and the $D$ parameters.}
The sixth term of ~\eqref{eq:F_123} contains
\begin{equation}
  T_6
  = 32\Bigl(
      \mathrm{Re}\{C_1^{*} \bar D_{12}^{*}\bar D_{13}^{*}D_{23}\}
      + \mathrm{Re}\{C_1^{*}D_{12}D_{13}D_{23}^{*}\}
    \Bigr).
\end{equation}
Using ~\eqref{eq:basic-CS},~\eqref{eq:D-bounds}, and the exact intensity expression \eqref{eq:C-exact-intensity}, we find
\begin{align}
  |T_6|
  &\le 32|C_1|\bigl(
          |\bar D_{12}|\,|\bar D_{13}|\,|D_{23}|
          + |D_{12}|\,|D_{13}|\,|D_{23}|
        \bigr)
  \nonumber\\
  &\le 64 |C_1|\,
      \bigl(\Sigma_{12}\Sigma_{13}\Sigma_{23}\bigr)^{1/2}
  \nonumber\\
  &= 64
      \sqrt{\langle W_1^{2}\rangle - 2\langle W_1\rangle^{2}}\,
      \bigl(\Sigma_{12}\Sigma_{13}\Sigma_{23}\bigr)^{1/2}.
  \label{eq:T6-tight-intensity}
\end{align}

\paragraph{Term involving $C_2$ and the $D$ parameters.}

The last term in ~\eqref{eq:F_123} is
\begin{equation}
  T_7
  = -64\,\mathrm{Re}\{C_2 D_{13}D_{12}^{*}D_{23}^{*}\}.
\end{equation}
Again, by ~\eqref{eq:basic-CS}, ~\eqref{eq:D-bounds}, and
~\eqref{eq:C-exact-intensity},
\begin{align}
  |T_7|
  &\le 64 |C_2|\,|D_{13}|\,|D_{12}|\,|D_{23}|
  \nonumber\\
  &\le 64 |C_2|\,
      \bigl(\Sigma_{12}\Sigma_{13}\Sigma_{23}\bigr)^{1/2}
  \nonumber\\
  &= 64
      \sqrt{\langle W_2^{2}\rangle - 2\langle W_2\rangle^{2}}\,
      \bigl(\Sigma_{12}\Sigma_{13}\Sigma_{23}\bigr)^{1/2}.
  \label{eq:T7-tight-intensity}
\end{align}

\paragraph{Collecting the intensity-only bound for $R_{123}$.}

Using the triangle inequality and
~\eqref{eq:T1-tight-intensity}–\eqref{eq:T7-tight-intensity}, we obtain
\begin{align}
  |\tilde F_{123}|
  &\le 16\bigl|B_1^{2}+B_1- |C_1|^{2}\bigr|\,\Sigma_{23}
       + 16\,\Sigma_{12}\Sigma_{13}
  \nonumber\\
  &\quad
       + \Bigl[
            16
            + 32|B_1 + B_2 - B_3|
            + 16(2B_1+1)
         \Bigr]
         \bigl(\Sigma_{12}\Sigma_{13}\Sigma_{23}\bigr)^{1/2}
  \nonumber\\
  &\quad
       + 64
         \sqrt{\langle W_1^{2}\rangle - 2\langle W_1\rangle^{2}}\,
         \bigl(\Sigma_{12}\Sigma_{13}\Sigma_{23}\bigr)^{1/2}
  \nonumber\\
  &\quad
       + 64
         \sqrt{\langle W_2^{2}\rangle - 2\langle W_2\rangle^{2}}\,
         \bigl(\Sigma_{12}\Sigma_{13}\Sigma_{23}\bigr)^{1/2}.
  \label{eq:R123-tight-intensity}
\end{align}
This bound depends only on the intensity correlation moments
$\langle W_j\rangle = B_j$, $\langle W_j^{2}\rangle$, and
$\langle W_j W_k\rangle$ via $\Sigma_{jk}$.

The full triple-beam contribution in ~\eqref{eq:ppt criterion} is obtained by adding $\mathrm{addt3}$, i.e.\ by summing expressions of the same form as~\eqref{eq:R123-tight-intensity} over all permutations of $(1,2,3)$.

\bibliography{ref}